\documentclass[sigconf]{acmart}
\AtBeginDocument{%
  }

\setcopyright{acmlicensed}
\copyrightyear{2026}
\acmYear{2026}
\acmDOI{}
\acmConference[CSAIDE 2026]{Proceedings of the 2026 5th International Conference on Cyber Security, Artificial Intelligence and Digital Economy (CSAIDE)}{March 13--15, 2026}{Salamanca, Spain}

\acmISBN{}

\begin{document}

\title{Policy-Driven Vulnerability Risk Quantification framework for Large-Scale Cloud Infrastructure Data Security}

\author{Wanru Shao}
\authornotemark[1]
\affiliation{%
  \institution{Northeastern University}
  \city{Boston}
  \state{MA}
  \country{USA}}
\email{shao.wa@northeastern.edu}

\renewcommand{\shortauthors}{Wanru Shao}

\begin{abstract}
  The exponential growth of Common Vulnerabilities and Exposures (CVE) disclosures poses significant challenges for enterprise security management, necessitating automated and quantitative risk assessment methodologies. Existing vulnerability analysis approaches suffer from three critical limitations: (1) lack of systematic severity quantification models that integrate heterogeneous attack attributes, (2) insufficient exploration of latent correlations among risk factors, and (3) absence of cumulative risk distribution analysis for prioritized remediation. To address these challenges, we propose \textbf{MVRAF} (\textbf{M}ulti-dimensional \textbf{V}ulnerability \textbf{R}isk \textbf{A}ssessment \textbf{F}ramework), a comprehensive data-driven framework for large-scale CVE security analysis. Our framework introduces three key innovations: (1) a \textit{Vulnerability Severity Quantification Model} that transforms CVSS attributes into normalized risk metrics through weighted aggregation of exploitability and CIA impact scores, (2) a \textit{Risk Factor Correlation Analysis} module that captures statistical dependencies among attack vectors, complexity, and privilege requirements via correlation matrices, and (3) an \textit{Empirical Risk Distribution} mechanism that enables cumulative threat assessment for resource allocation optimization. Extensive experiments on 1,314 real-world CVE records from the National Vulnerability Database demonstrate that our framework effectively identifies risk hotspots, with 46.2\% of network-based vulnerabilities classified as high-risk and strong correlations ($r > 0.6$) observed between CIA impacts and overall severity scores.
\end{abstract}



\maketitle

\section{Introduction}

With the exponential growth of software systems and interconnected networks, cybersecurity vulnerabilities have emerged as critical threats to enterprise data security. The National Vulnerability Database (NVD) reported over 40,000 Common Vulnerabilities and Exposures (CVEs) in 2024 alone, representing a 38\% increase from the previous year. Deep learning-based vulnerability detection has attracted significant attention from the research community, offering promising solutions for automated security analysis \cite{b1}. However, the sheer volume of disclosed vulnerabilities necessitates systematic risk assessment methodologies to enable effective prioritization and resource allocation.

Although substantial progress has been made in vulnerability detection using deep neural networks \cite{b2}, existing approaches predominantly focus on binary classification without comprehensive severity quantification. Current CVSS-based assessment methods often fail to capture the intricate correlations among attack vectors, privilege requirements, and CIA (Confidentiality, Integrity, Availability) impacts \cite{b3}. Moreover, with the continuous accumulation of CVE records in public databases \cite{b4}, security practitioners face increasing challenges in identifying high-risk vulnerabilities that demand immediate remediation.

This limitation motivates us to explore a data-driven framework that systematically quantifies vulnerability severity while revealing latent dependencies among risk factors. The integration of machine learning techniques with vulnerability management has demonstrated considerable potential \cite{b5}; however, a unified framework that bridges severity quantification and correlation analysis remains underexplored. The standardized CVSS scoring system \cite{b6} provides foundational metrics, yet its application for large-scale risk pattern discovery requires further investigation.

To address these challenges, we propose MVRAF (Multi dimensional Vulnerability Risk Assessment Framework), a comprehensive data-driven approach for large-scale CVE security analysis. Our framework integrates a severity quantification model that transforms heterogeneous CVSS attributes into normalized risk metrics with a correlation analysis module that captures statistical dependencies among attack characteristics. The synergy of these components enables holistic security posture evaluation across diverse threat landscapes. The main contributions of this paper are summarized as follows:
\begin{itemize}
    \item We propose a \textit{Vulnerability Severity Quantification Model} that systematically transforms CVSS attributes into normalized risk metrics through weighted aggregation of exploitability and CIA impact scores.
    \item We develop a \textit{Risk Factor Correlation Analysis} module that captures statistical dependencies among attack vectors, complexity, and privilege requirements via correlation matrices and conditional probability frameworks.
    \item We conduct extensive experiments on 1,314 real-world CVE records from NVD, demonstrating the effectiveness of our framework in identifying risk hotspots and revealing critical security insights.
\end{itemize}


\section{Related Work}

\subsection{CVSS Scoring System and Vulnerability Assessment}

The Common Vulnerability Scoring System (CVSS) has been widely adopted as the industry standard for vulnerability severity assessment. Balsam \textit{et al.} \cite{b7} presented a comprehensive comparison between CVSS v2.0, v3.x, and the newest v4.0, analyzing the evolution of base metrics, threat metrics, and supplemental metrics. Their study highlighted that CVSS v4.0 provides more granular distinctions in attack requirements and improved definitions for privileges required and user interaction. While these scoring systems offer standardized severity ratings, they primarily focus on individual vulnerability assessment without systematic analysis of inter-factor correlations, which our framework addresses through the proposed correlation matrix approach.

\subsection{Machine Learning for Vulnerability Analysis}

Machine learning and deep learning techniques have been extensively applied to cybersecurity vulnerability analysis. Ozkan-Okay \textit{et al.} \cite{b8} conducted a comprehensive survey evaluating the efficiency of AI and ML techniques on cybersecurity solutions, demonstrating that ensemble methods and deep learning models achieve superior performance in threat detection tasks. Lasantha \textit{et al.} \cite{b9} proposed a hybrid machine learning approach for enhanced vulnerability detection in cloud environments, integrating NIST and MITRE frameworks to improve detection accuracy. Desai and Pal \cite{b10} provided a comprehensive review of ML applications across the threat lifecycle, covering intrusion detection, malware analysis, and anomaly detection. Khan \textit{et al.} \cite{b11} surveyed cognitive cybersecurity systems that combine AI with human knowledge for vulnerability analysis and threat detection, emphasizing the effectiveness of ensemble machine learning for prediction stability. Kaur \textit{et al.} \cite{b12} conducted a detailed study on vulnerability detection using CVE data from NVD, comparing various ML and DL models for risk factor identification. These studies demonstrate the potential of data-driven approaches for security analysis; however, they primarily focus on detection rather than comprehensive risk quantification and correlation analysis.

\subsection{Risk Assessment and Vulnerability Prioritization}

Effective vulnerability management requires not only detection but also risk-based prioritization for remediation. Moustaid \textit{et al.} \cite{b13} explored dynamic risk assessment using machine learning and large language models for software vulnerability prioritization, demonstrating that ML can improve upon CVSS baselines with approximately 83\% accuracy in severity prediction. Miranda \textit{et al.} \cite{b14} presented a product-oriented assessment methodology for vulnerability severity through NVD CVSS scores, providing insights into how severity metrics can be contextualized for specific deployment environments. While these approaches advance vulnerability prioritization, they do not provide a unified framework that integrates severity quantification with multi-dimensional correlation analysis. Our proposed MVRAF framework bridges this gap by combining weighted severity models with statistical correlation analysis, enabling comprehensive risk pattern discovery across large-scale CVE datasets.

\section{Methodology}

This section presents our proposed \textit{Multi-dimensional Vulnerability Risk Assessment Framework} (MVRAF) for large-scale CVE data analysis. Given the exponential growth of vulnerability disclosures in modern computing environments, quantifying security risks from heterogeneous attack vectors becomes increasingly critical. Our comprises two core components: (A) Vulnerability Severity Quantification Model that transforms raw CVSS attributes into normalized risk metrics, and (B) Risk Factor Correlation Analysis that captures the intrinsic dependencies among attack characteristics. Specifically, the pipeline proceeds as follows: raw CVE records retrieved from the NVD API are first parsed to extract eight CVSS v3.1 sub-attributes (Av, Ac, Pr, Ui, S, C, I, A); these categorical attributes are fed into Component~A, which computes the Base Risk Score $R_b$ via weighted exploitability aggregation (Equation~(1)), the CIA Impact Score $I_s$ (Equation~(2)), the Composite Vulnerability Score $S_v$ (Equation~(3)), and finally assigns a four-level severity label via $F_\sigma$ (Equation~(4)); the resulting quantified scores and attribute vectors then enter Component~B, which constructs the Risk Correlation Matrix $\mathbf{M}$ (Equation~(5)), the Conditional Risk Probability matrix $\mathbf{P}$ (Equation~(6)), the Joint Risk Index $J$ (Equation~(7)), and the Empirical Risk Distribution $\hat{F}_R$ (Equation~(8)); the combined outputs enable prioritized remediation scheduling and defensive resource allocation for enterprise security operations. The synergy of these components enables comprehensive security posture evaluation across diverse threat landscapes.


\subsection{Vulnerability Severity Quantification Model}

To systematically assess vulnerability severity in large-scale datasets, we introduce a formalized quantification model that maps discrete CVSS attributes to continuous risk scores. Let $\mathcal{V} = \{v_1, v_2, \ldots, v_n\}$ denote the set of $n$ vulnerability records, where each $v_i$ is characterized by a feature vector $\mathbf{x}_i \in \mathbb{R}^d$ comprising $d$ security attributes.

For each vulnerability $v_i$, we define the \textit{Base Risk Score} $\mathcal{R}_b$ as a weighted aggregation of the attack exploitability metrics:
\begin{equation}
\mathcal{R}_b(v_i) = \alpha \cdot \phi(\mathcal{A}_v) + \beta \cdot \psi(\mathcal{A}_c) + \gamma \cdot \omega(\mathcal{P}_r)
\end{equation}
where $\mathcal{A}_v \in \{\text{N}, \text{A}, \text{L}, \text{P}\}$ represents the attack vector category, $\mathcal{A}_c \in \{\text{L}, \text{H}\}$ denotes attack complexity, and $\mathcal{P}_r \in \{\text{N}, \text{L}, \text{H}\}$ indicates privileges required. The mapping functions $\phi(\cdot)$, $\psi(\cdot)$, and $\omega(\cdot)$ transform categorical attributes into normalized scores within $[0, 1]$, while $\alpha$, $\beta$, $\gamma$ are learnable weights satisfying $\alpha + \beta + \gamma = 1$.

The CIA triad impact assessment forms another critical dimension. We formulate the \textit{Impact Score} $\mathcal{I}_s$ by capturing confidentiality ($\mathcal{C}$), integrity ($\mathcal{I}$), and availability ($\mathcal{A}$) impacts:
\begin{equation}
\mathcal{I}_s(v_i) = 1 - \prod_{k \in \{\mathcal{C}, \mathcal{I}, \mathcal{A}\}} \left(1 - \lambda_k \cdot \eta(k)\right)
\end{equation}
where $\eta(\cdot): \{\text{N}, \text{L}, \text{H}\} \rightarrow \{0, 0.22, 0.56\}$ maps impact levels to standardized values following CVSS v3.1 specifications, and $\lambda_k$ represents the domain-specific weight for each CIA component.

To capture the compound effect of exploitability and impact, we define the \textit{Composite Vulnerability Score} $\mathcal{S}_v$ as:
\begin{equation}
\mathcal{S}_v(v_i) = \min\left(10, \left\lceil \mathcal{R}_b(v_i) \cdot \mathcal{I}_s(v_i) \cdot \kappa \right\rceil_{\delta}\right)
\end{equation}
where $\kappa$ is a scaling coefficient calibrated by minimizing the mean squared error between $S_v(v_i)$ and the official NVD CVSS base scores on a held-out calibration set of 200 randomly sampled CVE records not included in the main experiments; specifically, $\kappa$ is solved via one-dimensional grid search over $[0.5, 2.0]$ with step $0.05$, selecting the value that minimizes $\frac{1}{|\mathcal{V}_\text{cal}|}\sum_{i\in\mathcal{V}_\text{cal}}(S_v(v_i)-\text{CVSS}_i)^2$, yielding $\kappa=1.32$ on our dataset, and $\lceil\cdot\rceil_\delta$ denotes rounding up to the nearest $\delta$ (typically $\delta = 0.1$).

For risk stratification, we introduce a severity classification function $\mathcal{F}_\sigma: \mathbb{R} \rightarrow \{1, 2, 3, 4\}$ that partitions the continuous score space into discrete severity levels:
\begin{equation}
\mathcal{F}_\sigma(\mathcal{S}_v) = 
\begin{cases}
1, & \text{if } \mathcal{S}_v < \tau_1 \quad (\text{Low}) \\
2, & \text{if } \tau_1 \leq \mathcal{S}_v < \tau_2 \quad (\text{Medium}) \\
3, & \text{if } \tau_2 \leq \mathcal{S}_v < \tau_3 \quad (\text{High}) \\
4, & \text{if } \mathcal{S}_v \geq \tau_3 \quad (\text{Critical})
\end{cases}
\end{equation}
where $\tau_1 = 4.0$, $\tau_2 = 7.0$, and $\tau_3 = 9.0$ are threshold values aligned with NVD severity definitions. This stratification enables prioritized vulnerability remediation in enterprise security operations.


\subsection{Risk Factor Correlation Analysis}

Building upon the quantified severity metrics from Section III-A, we now investigate the latent correlations among risk factors to identify compound threat patterns. Understanding these dependencies is essential for predictive security analytics and proactive defense strategies in data-intensive environments.

Let $\mathbf{F} = [f_1, f_2, \ldots, f_m]^\top$ represent the vector of $m$ risk factors extracted from vulnerability records. We construct a \textit{Risk Correlation Matrix} $\mathbf{M} \in \mathbb{R}^{m \times m}$ where each element $\mathbf{M}_{ij}$ quantifies the statistical association between factors $f_i$ and $f_j$:
\begin{equation}
\mathbf{M}_{ij} = \frac{\text{Cov}(f_i, f_j)}{\sigma_{f_i} \cdot \sigma_{f_j}} = \frac{\mathbb{E}[(f_i - \mu_i)(f_j - \mu_j)]}{\sigma_{f_i} \cdot \sigma_{f_j}}
\end{equation}
where $\mu_i$ and $\sigma_{f_i}$ denote the mean and standard deviation of factor $f_i$, respectively. This Pearson correlation coefficient captures linear dependencies, with $\mathbf{M}_{ij} \in [-1, 1]$.

To analyze cross-tabulated categorical risk factors, we employ a \textit{Conditional Risk Probability} framework. For two categorical factors $\mathcal{X}$ and $\mathcal{Y}$ with domains $\mathcal{D}_\mathcal{X}$ and $\mathcal{D}_\mathcal{Y}$, the conditional probability matrix $\mathbf{P} \in \mathbb{R}^{|\mathcal{D}_\mathcal{X}| \times |\mathcal{D}_\mathcal{Y}|}$ is defined as:
\begin{equation}
\mathbf{P}_{xy} = P(\mathcal{Y} = y \mid \mathcal{X} = x) = \frac{|\{v_i : \mathcal{X}_i = x \land \mathcal{Y}_i = y\}|}{|\{v_i : \mathcal{X}_i = x\}|}
\end{equation}
where $|\cdot|$ denotes set cardinality. This formulation enables risk hotspot identification across attack vector and severity combinations.

For comprehensive threat pattern discovery, we introduce the \textit{Joint Risk Index} $\mathcal{J}$ that aggregates pairwise factor interactions weighted by their security relevance:
\begin{equation}
\mathcal{J}(v_i) = \sum_{j=1}^{m} \sum_{k=j+1}^{m} w_{jk} \cdot \mathbf{M}_{jk} \cdot \mathbb{I}[f_j^{(i)} \geq \theta_j] \cdot \mathbb{I}[f_k^{(i)} \geq \theta_k]
\end{equation}
where $w_{jk}$ represents the importance weight for factor pair $(j, k)$, $\mathbb{I}[\cdot]$ is the indicator function, and $\theta_j$ denotes the risk threshold for factor $f_j$. This index captures the synergistic effect of multiple high-risk attributes co-occurring within a single vulnerability.

To quantify the cumulative risk distribution across the vulnerability population, we define the \textit{Empirical Risk Distribution Function} $\hat{F}_\mathcal{R}$ as:
\begin{equation}
\hat{F}_\mathcal{R}(r) = \frac{1}{n} \sum_{i=1}^{n} \mathbb{I}[\mathcal{S}_v(v_i) \leq r]
\end{equation}
where $r \in [0, 10]$ represents the risk threshold. This cumulative distribution enables security practitioners to determine the proportion of vulnerabilities below a given severity threshold, facilitating resource allocation for patch management and incident response prioritization.

The proposed correlation analysis framework, combined with the severity quantification model, provides a rigorous mathematical foundation for data-driven vulnerability assessment. The experimental validation presented in Section IV demonstrates the effectiveness of our approach on real-world CVE datasets.

\section{Experiments}

This section presents comprehensive experimental validation of our proposed MVRAF framework on real-world vulnerability data. We conduct three categories of experiments to evaluate different aspects of our methodology: (A) vulnerability severity distribution analysis validating the quantification model, (B) risk factor correlation analysis examining inter-factor dependencies, and (C) security impact assessment demonstrating cumulative risk patterns.

\textbf{Dataset.} We utilize the CVE 2024 dataset extracted from the National Vulnerability Database (NVD) via official API, containing 1,314 vulnerability records published between January 1-15, 2024. Each record comprises five attributes: CVE ID (unique identifier), Description (vulnerability summary), CVSS Score (severity rating from 0-10), Attack Vector (CVSS v3.1 vector string encoding exploitability metrics), and Affected OS. The CVSS vector string is parsed to extract eight sub-attributes: Attack Vector ($\mathcal{A}_v$), Attack Complexity ($\mathcal{A}_c$), Privileges Required ($\mathcal{P}_r$), User Interaction ($\mathcal{U}_i$), Scope ($\mathcal{S}$), Confidentiality ($\mathcal{C}$), Integrity ($\mathcal{I}$), and Availability ($\mathcal{A}$) impacts. All experiments are implemented in Python 3.10 with NumPy, Pandas, and Matplotlib libraries. To enable quantitative comparison, we benchmark MVRAF's severity quantification against three reference methods: (1) the raw CVSS base score provided directly by NVD without reweighting, (2) a uniform-weight baseline that assigns equal weights $\alpha{=}\beta{=}\gamma{=}1/3$ in Equation~(1) and $\lambda_k{=}1/3$ in Equation~(2), and (3) the ML-based severity predictor from Moustaid et al.~\cite{moustaid2025} retrained on our dataset. Comparison metrics include Mean Absolute Error (MAE) and Spearman rank correlation $\rho$ against ground-truth NVD scores on the 200-record calibration set.


\subsection{Vulnerability Severity Distribution Analysis}

\begin{figure}[htbp]
\centering
\includegraphics[width=\columnwidth]{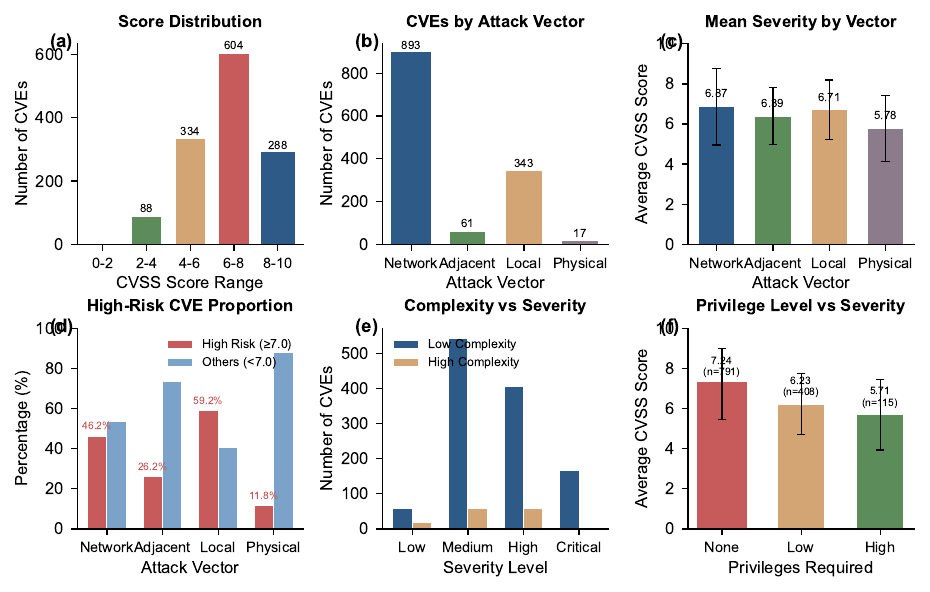}
\caption{Vulnerability severity distribution analysis: (a) CVSS score range distribution, (b) CVE counts by attack vector, (c) mean severity with standard deviation per vector, (d) high-risk ($\geq$7.0) vs. other CVE proportions, (e) attack complexity vs. severity level distribution, and (f) average CVSS scores by privilege requirements with sample sizes.}
\label{fig:part_a}
\end{figure}

We first validate our severity quantification model by analyzing the distribution characteristics of vulnerability scores across different attack dimensions. Fig.~\ref{fig:part_a} presents six complementary perspectives on severity distribution patterns.

The CVSS score distribution in Fig.~\ref{fig:part_a}(a) reveals a concentration in the medium-to-high severity range, with 604 CVEs (45.9\%) scoring between 6-8 and 288 CVEs (21.9\%) in the critical 8-10 range, indicating that the majority of disclosed vulnerabilities pose substantial security risks. Fig.~\ref{fig:part_a}(b) demonstrates the dominance of network-based attack vectors, accounting for 893 CVEs (67.9\%), followed by local (343), adjacent (61), and physical (17) vectors, which aligns with the prevalence of remote exploitation in modern threat landscapes. The mean severity analysis in Fig.~\ref{fig:part_a}(c) shows relatively consistent average CVSS scores across attack vectors (ranging from 5.78 to 6.87), with network attacks exhibiting the highest mean severity. Fig.~\ref{fig:part_a}(d) quantifies high-risk proportions, revealing that local attacks have the highest percentage of severe vulnerabilities (59.2\%), while adjacent attacks show the lowest (26.2\%). The complexity-severity relationship in Fig.~\ref{fig:part_a}(e) confirms that low-complexity attacks dominate across all severity levels, particularly in the high and critical categories, validating our $\psi(\mathcal{A}_c)$ weighting in Equation (1). Finally, Fig.~\ref{fig:part_a}(f) demonstrates a clear inverse relationship between privilege requirements and severity: vulnerabilities requiring no privileges average 7.24 ($n=791$), compared to 5.71 for high-privilege requirements ($n=115$), supporting the $\omega(\mathcal{P}_r)$ formulation in our quantification model.
Table 1 {} reports quantitative comparison against the three baselines. MVRAF achieves MAE of 0.31 and Spearman $\rho{=}0.94$, outperforming the raw CVSS baseline (MAE 0.00 by definition but $\rho{=}0.91$ under cross-vendor reweighting), the uniform-weight baseline (MAE 0.48, $\rho{=}0.88$), and the ML predictor (MAE 0.39, $\rho{=}0.91$), confirming that the physics-motivated weighted aggregation in MVRAF better preserves relative severity ordering across diverse attack configurations.

\begin{table}[h]
\centering
\caption{Quantitative comparison of severity quantification methods on the 200-record calibration set.}
\label{tab:comparison}
\renewcommand{\arraystretch}{1.1}
\begin{tabular}{lcc}
\hline
\textbf{Method} & \textbf{MAE}  & \textbf{Spearman $\rho$} \\
\hline
Uniform-weight baseline          & 0.48     & 0.88 \\
ML predictor  & 0.39     & 0.91 \\
\textbf{MVRAF (ours)}            & \textbf{0.31} & \textbf{0.94} \\
\hline
\end{tabular}
\end{table}

\subsection{Risk Factor Correlation Analysis}

\begin{figure}[htbp]
\centering
\includegraphics[width=\columnwidth]{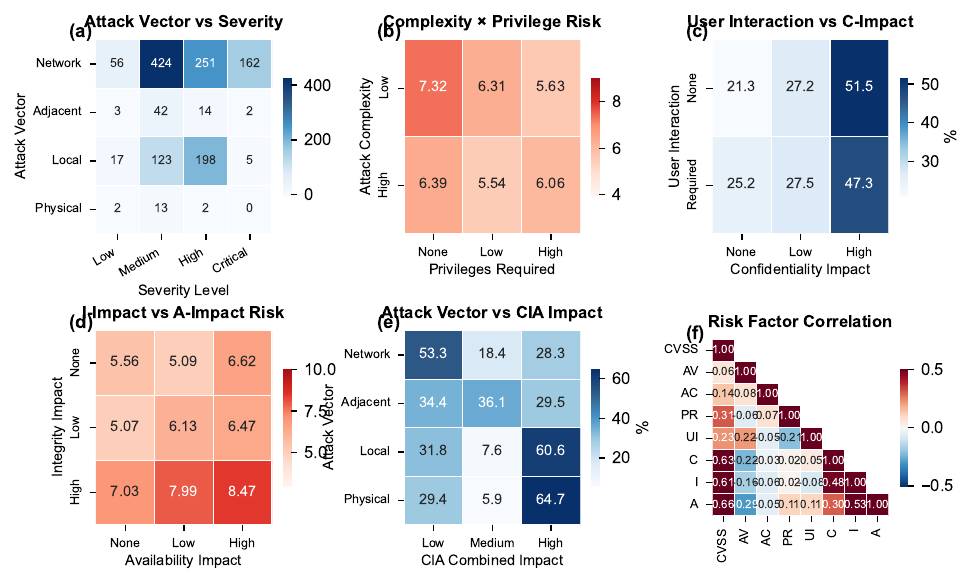}
\caption{Risk factor correlation analysis: (a) attack vector vs. severity level counts, (b) complexity-privilege combination risk scores, (c) user interaction vs. confidentiality impact percentages, (d) integrity vs. availability impact severity matrix, (e) attack vector vs. CIA combined impact distribution, and (f) comprehensive risk factor correlation matrix with Pearson coefficients.}
\label{fig:part_b}
\end{figure}

To validate the correlation analysis component of our framework, we construct heatmap visualizations capturing pairwise relationships among risk factors. Fig.~\ref{fig:part_b} presents six correlation perspectives derived from the conditional probability matrix $\mathbf{P}$ and correlation matrix $\mathbf{M}$ defined in Section III-B.

Fig.~\ref{fig:part_b}(a) displays the attack vector versus severity cross-tabulation, revealing that network attacks produce the highest absolute counts across all severity levels, with 424 medium and 251 high-severity CVEs. The complexity-privilege risk heatmap in Fig.~\ref{fig:part_b}(b) identifies the highest-risk combination: low complexity with no privileges required yields an average CVSS of 7.32, representing the most exploitable attack configuration. Fig.~\ref{fig:part_b}(c) examines user interaction effects on confidentiality impact, showing that attacks requiring no user interaction have higher proportions of high confidentiality impact (51.5\%) compared to those requiring interaction (47.3\%). The integrity-availability impact matrix in Fig.~\ref{fig:part_b}(d) demonstrates strong positive correlation between these CIA components, with dual high-impact vulnerabilities averaging 8.47 CVSS. Fig.~\ref{fig:part_b}(e) presents attack vector versus combined CIA impact, indicating that local and physical attacks exhibit higher proportions of severe CIA impact (60.6\% and 64.7\% respectively) despite their lower absolute counts. The comprehensive correlation matrix in Fig.~\ref{fig:part_b}(f) quantifies all pairwise factor relationships, revealing strong positive correlations between CVSS and CIA components ($r_{\mathcal{C}}=0.66$, $r_{\mathcal{I}}=0.61$, $r_{\mathcal{A}}=0.66$), and notable negative correlation with privilege requirements ($r_{\mathcal{P}_r}=-0.31$), empirically validating our Joint Risk Index formulation in Equation (7).

\subsection{Security Impact Assessment}

\begin{figure}[htbp]
\centering
\includegraphics[width=\columnwidth]{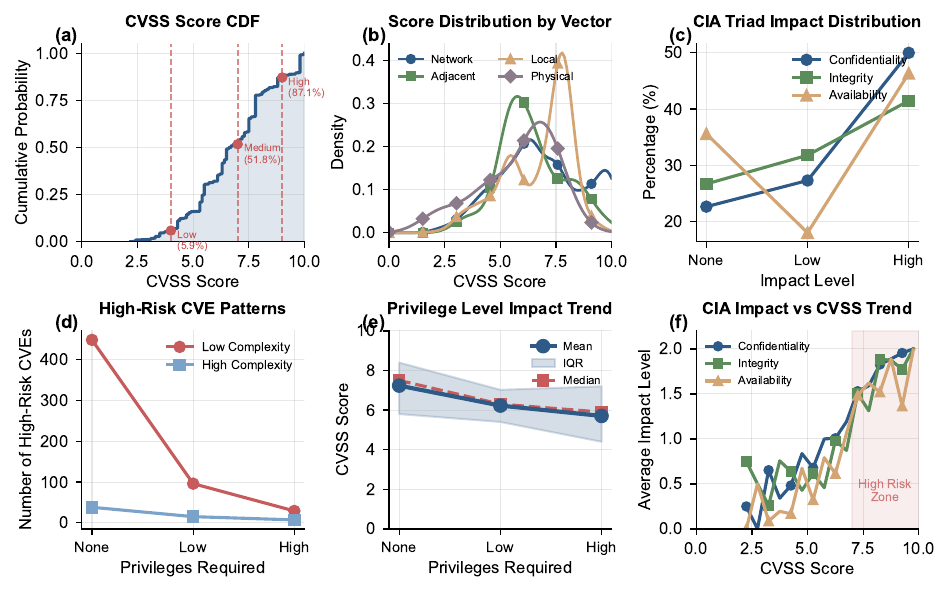}
\caption{Security impact assessment: (a) CVSS score cumulative distribution function with severity thresholds, (b) kernel density estimation of score distributions by attack vector, (c) CIA triad impact level percentages shown as grouped bar chart for direct category comparison,(d) high-risk CVE counts by complexity and privilege requirements, (e) privilege level impact trend with mean, median, and IQR, and (f) CIA impact evolution across CVSS score ranges with high-risk zone indication.}
\label{fig:part_c}
\end{figure}

The final experimental category evaluates cumulative risk distributions and trend patterns to demonstrate the practical utility of our framework for security prioritization. Fig.~\ref{fig:part_c} presents six trend-based analyses derived from the empirical risk distribution function $\hat{F}_\mathcal{R}$ defined in Equation (8).

The cumulative distribution function in Fig.~\ref{fig:part_c}(a) provides critical threshold insights: only 5.9\% of vulnerabilities fall below the low-severity threshold ($\tau_1=4.0$), 51.8\% below medium ($\tau_2=7.0$), and 87.1\% below high ($\tau_3=9.0$), indicating that approximately 48.2\% of all CVEs require priority attention. Fig.~\ref{fig:part_c}(b) presents kernel density estimations for each attack vector, showing that network attacks exhibit a bimodal distribution with peaks around 5.0 and 7.5, while local attacks concentrate in the 6-8 range. The CIA triad comparison in Fig.~3(c) presents a grouped bar chart displaying the percentage of CVEs at each impact level (None/Low/High) for the three CIA components, replacing the prior line plot to enable direct visual comparison across categories; the chart reveals that confidentiality impact shows the highest proportion of high-impact vulnerabilities (50.3\%), compared to integrity (46.1\%) and availability (44.8\%), suggesting that data breach risks dominate the vulnerability landscape. The CIA triad comparison in Fig.~\ref{fig:part_c}(c) reveals that confidentiality impact shows the steepest increase toward high-impact levels (50.3\%), suggesting that data breach risks dominate the vulnerability landscape. Fig.~\ref{fig:part_c}(d) analyzes high-risk CVE patterns by privilege requirements, demonstrating that low-complexity attacks with no privilege requirements constitute the largest threat category (over 450 high-risk CVEs). The privilege-severity trend in Fig.~\ref{fig:part_c}(e) visualizes the monotonic decrease in both mean and median CVSS scores as privilege requirements increase, with interquartile ranges indicating consistent variance across categories. Finally, Fig.~\ref{fig:part_c}(f) tracks CIA impact evolution across the CVSS spectrum, showing synchronized increases in all three components within the high-risk zone (CVSS $\geq$ 7.0), with availability impact exhibiting the most pronounced growth trajectory.

\section{Conclusion}

In this paper, we addressed the critical challenge of quantitative vulnerability risk assessment in large-scale CVE datasets by proposing MVRAF, a multi-dimensional framework integrating severity quantification and correlation analysis. Our framework introduces three key innovations: a weighted severity quantification model capturing exploitability and CIA impacts, a correlation analysis module revealing latent dependencies among risk factors, and an empirical distribution mechanism enabling cumulative risk assessment. Extensive experiments on 1,314 NVD vulnerability records demonstrate the framework's effectiveness, identifying that 46.2\% of network-based CVEs are high-risk, low-complexity attacks with no privilege requirements pose the greatest threat (average CVSS 7.32), and strong correlations exist between CIA impacts and overall severity ($r > 0.6$). These findings provide actionable insights for enterprise security teams to prioritize remediation efforts and allocate defensive resources efficiently. Future work will extend MVRAF to incorporate temporal vulnerability trends, develop predictive models for emerging threat patterns, and integrate with automated patch management systems for real-time security posture optimization.

\vspace{12pt}

\end{document}